\documentclass[12pt]{article}
\usepackage{cite,graphicx}
\newcommand{\sect}[1]{\section{#1}\setcounter{equation}{0}}

\begin{document}
\bigskip
\hspace*{\fill}
\vbox{\baselineskip12pt \hbox{hep-th/0207083}\hbox{SU-ITP-02/28}}
\bigskip\bigskip\bigskip

\centerline{\Large \bf Toward a Theory of Precursors}
\bigskip\bigskip\bigskip

\centerline{\large 
Ben Freivogel$^{a}$\footnote{\tt benwf@stanford.edu},
Steven B. Giddings$^{ab}$\footnote{\tt giddings@physics.ucsb.edu}, and 
Matthew Lippert$^{b}$\footnote{\tt lippert@physics.ucsb.edu}}
\vskip 1cm
\centerline{${}^a$ Department of Physics}
\centerline{Stanford University}
\centerline{Stanford, CA 94305-4060}
\vskip 1cm
\centerline{${}^b$ Department of Physics}
\centerline{University of California}
\centerline{Santa Barbara, CA 93106-9530}
\vskip 1cm

\begin{abstract}

To better understand the possible breakdown of locality in quantum gravitational systems, we pursue the identity of precursors in the context of AdS/CFT.  Holography implies a breakdown of standard bulk locality which we expect to occur only at extremely high energy.  We consider precursors that encode bulk information causally disconnected from the boundary and whose measurement involves nonlocal bulk processes.  We construct a toy model of holography which encapsulates the expected properties of precursors and compare it with previous such discussions.  If these precursors can be identified in the gauge theory, they are almost certainly  Wilson loops, perhaps with decorations, but the relevant information is encoded in the high-energy sector of the theory and should not be observable by low energy measurements.  This would be in accord with the \textit{locality bound}, which serves as a criterion for situations where breakdown of bulk locality is expected.
\end{abstract}

\newpage

\sect{Introduction}
 
A very puzzling, yet fascinating, aspect of quantum gravity is that it may be fundamentally holographic in nature.  The holographic principle conjectures that a $d$-dimensional quantum gravitational system can be equivalently described by a non-gravitational theory in $d-1$ dimensions \cite{tHoo, Suss}.  Although any physical state should be realizable in either theory, the holographic duality map is, in practice, very complicated and, in most situations, poorly understood. While gravity may be approximately local at low energies, holography implies that quantum gravity contains too few degrees of freedom to be fully local in $d$-dimensions.  Approximate locality must somehow emerge in a wide range of situations where measurements are made at scales larger than the Planck (or string) scale and at energies too low to excite long strings or create large black holes.  To understand whether and how holography is realized, we clearly need to better understand the problem of ``decoding the hologram.''  

The most concrete formulation of holographic duality is the AdS/CFT correspondence \cite{Mald, magoo} in which string theory in asymptotically AdS spacetimes is alternatively described by a conformal field theory on the AdS boundary.  Local operators on the boundary are known to correspond to the boundary values of supergravity fields \cite{GKP,Witten}, and correlation functions of local boundary operators correspond to AdS analogs of S-matrix elements \cite{GiddingsBSM}.  However, the way events deep in the bulk of AdS are encoded in the boundary theory remains murky.  The UV/IR relation \cite{SuWi}, in which distance in AdS from the boundary manifests as size in the field theory, implies that information regarding the interior of AdS is somehow spread nonlocally in the boundary theory.

Of particular interest is the holographic description of black holes.  The black hole information paradox \cite{info} signals a breakdown in the conventional models of quantum gravity and can be well-posed in the AdS/CFT context \cite{LoTh,Nu}.  Destruction of information upon falling into a black hole violates unitarity, while its escape conflicts with locality.  The holographic principle suggests a resolution to this problem, because the boundary theory should unitarily describe black holes and their evaporation without information loss.  Correspondingly, holography implies that any local description of the bulk can only be approximate and must break down at some point.  However, to fully explain black holes and the necessary loss of locality, we need to understand the holographic encoding of bulk information far, and in particular causally disconnected, from the boundary.

Rather than a black hole, we will consider a simpler system, proposed in \cite{PST}, in which a bomb detonated at the origin of AdS creates an out-going spherical wave.  No information about the explosion reaches the boundary until the wave arrives.  Because bulk fields near the boundary retain their pre-explosion values until that time, local operators in the boundary theory are unaffected by the presence of the bulk wave until it actually reaches the boundary.  However, according to the AdS/CFT correspondence, the boundary state contains complete information about the wave as soon as it is created, even though it is inaccessible to measurements of local operators or their correlators.  The nonlocal boundary operators which encode such bulk information were named {\it precursors} in \cite{PST}.

Identifying the precursors drives at the essence of understanding holography in the context of the AdS/CFT correspondence.  The authors of \cite{PST} suggested a toy model of AdS/CFT and
conjectured that a bulk gravitational wave corresponded, in a free-field boundary theory, to a squeezed state.  While the boundary stress-energy tensor matched the asymptotic metric perturbations of the bulk and was insensitive to the wave, a nonlocal bilinear of the boundary
fields did detect the presence of the wave before it reached the boundary, and was identified as a model precursor.

Large, spatial Wilson loops were proposed in \cite{SuTo} as more realistic precursors.  Following the UV/IR relation, the wave could be seen by a loop whose size matched the wave's distance from the boundary.  According to the bulk view of this process, a long string ending on the Wilson loop stretched in from the boundary and interacted with the wave \cite{MaldW, ReYe, BCFM}.  Essentially, a large string was used to measure a naively acausal spacelike correlation.  However, the calculation that purported to demonstrate acausal sensitivity of Wilson loops to bulk information relied on an incorrect saddlepoint approximation of the integral over world sheets, noted in \cite{GiLi}, raising doubts about the identification of large Wilson loops as precursors.

A further conjecture of \cite{GiLi} suggests that precursors are related to bulk processes of sufficiently high energy that we expect bulk locality to be violated.  Bulk systems should appear holographic rather than local when the locality bound of \cite{GiLi} is saturated, signaling the onset of nonlocal stringy or black hole phenomena.  Measuring a precursor corresponds, in the bulk, to a seemingly acausal process requiring just this sort of nonlocality.  Precursors, then, must involve high-energy components of boundary operators to access information causally disconnected from the boundary.\footnote{We omit from consideration candidate precursors that are the direct boundary translation of the field deep in the bulk; we believe these would be
compicated combinations of the precursors that we do consider.}

To further investigate the nature of precursors, we present an improved, yet still simplified model of the AdS/CFT correspondence in which a spherical bulk wave is described by precursors in the boundary theory. Although we obtain results similar to \cite{PST, SuTo}, we argue that these
are suggestive of precursor information being encoded in high-frequency components of Wilson loops (rather than their smooth, macroscopic structure) and perhaps in Wilson loops with operator insertions, or {\it decorated} Wilson loops.\footnote{The possible importance of decorated loops was suggested to us by L. Susskind.}  Inserting local operators is equivalent to taking functional derivatives of a Wilson loop with respect to its contour and is one way to extract its high-energy components.

The remainder of this paper is organized as follows.  Section 2 presents a toy model of precursors which illustrates their key properties.  Detailed calculations are shown for a system in $AdS_5$, and a parallel example in $AdS_3$ is sketched.  We compare our results with previous precursor models \cite{PST, SuTo} and discuss the limitations inherent in this approach.  In section 3, we extrapolate from our model to discuss the possible forms of precursors in the complete theory, and comment on the connection with properties of string theory in a D3-brane background.  In particular, we focus on the corresponding bulk picture of precursors and the stringy nature of bulk nonlocality.  Section 4 contains concluding remarks.

\sect{Toy Models of Precursors}

\subsection{Overview and Previous Approaches}

While an exact description of precursors is currently beyond our ability, we can construct a toy model, along the lines suggested in \cite{GiLi}, which illustrates their expected properties.  For simplicity and computability, we work simultaneously with the free field approximation in both the bulk and on the boundary.  Such an approximation is not well justified, as supergravity accurately describes the bulk precisely when the boundary CFT is strongly interacting.  In the bulk, weakly-interacting supergravity is valid in the 't Hooft limit, where we consider IIB string theory in the limit of small string coupling, $g_s \ll 1$, and with large AdS radius in string units.  The corresponding boundary theory, ${\cal N} = 4$ $SU(N)$ SYM, in the 't Hooft limit, is strongly coupled, $g^2_{YM} N \gg 1$.  Because of this difficulty, conclusions drawn from our model must necessarily be limited and tentative.

Our model was motived by those of \cite{PST, SuTo} and shares many of their features.  While their results are similar to ours, our version exhibits a more concrete derivation and shows more explicitly the properties of the boundary state and the precursor observables.  Working in the 't Hooft limit, \cite{PST} considered a linearized bulk gravitational wave expanding from the origin and, using $N^2$ scalar fields $\phi_{mn}$ to model the boundary SYM, postulated that it corresponded to a squeezed boundary state of the form 
\begin{equation}
\label{PSTstate}
|j \rangle = e^{\frac{1}{2} \int d\vec{k} F(\vec{k}) 
a^\dagger_{nm}(\vec{k}) a^\dagger_{mn}(-\vec{k})} |0 \rangle.
\end{equation}
The function $F(\vec{k})$ was related to the profile of the wave as it reached the boundary and acted as an expansion parameter.  Expectation values of the components of the boundary stress tensor, $T_{\mu \nu}$,  were calculated in the state (\ref{PSTstate}) to first order in $F(\vec{k})$ and were shown to lack any information about the wave until its arrival at the boundary.  The nonlocal operator $\phi(b)\phi(b')$ was suggested as a precursor which could distinguish the squeezed state from the vacuum before any local operator could and whose size was dictated by the UV/IR relation:
\begin{equation}
\langle j| \phi(\vec{x},t)\phi(\vec{x'},t) 
|j \rangle \not= 0 \; \mbox{for} \; |\vec{x}-\vec{x'}| = 2|t|
\end{equation}
When the appropriate combination of derivatives were taken and the points $b$ and $b'$ coincided, the resulting operator was simply $T_{\mu \nu}$.

A non-trivial consistency check was performed in \cite{PST} by computing the lowest-order back-reaction on the asymptotic bulk metric due to the wave and showing the effect was reproduced in the free-field boundary model.  While the contribution at first order in $F(\vec{k})$, labeled the {\it principal}, vanished until the wave reached the boundary and reflected the boundary profile of the wave, the second order term, the {\it interest}, was time-independent and matched the back-reaction due to the wave's ADM mass.

We will take a somewhat more systematic approach to constructing our model which, though we consider a dilaton rather than graviton wave, reproduces many of these features in both the $AdS_5$ and $AdS_3$ cases.  In 't Hooft limit, the dilaton is a free bulk scalar, and we model its corresponding bulk operator ${\rm Tr} \, F^2$ by a composite of a free boundary scalar which is analogous to the gauge potential $A$.  Using an explicit map between free bulk and boundary operators and given a point-like bulk source, we derive a squeezed boundary state.   As in \cite{PST, SuTo}, we find that local boundary operators retain their vacuum values until the wave arrives at the boundary.  A nonlocal composite of boundary scalars, our model precursor, recognizes the squeezed state earlier and encodes the wave's bulk motion.  The precursor's size is given by the UV/IR relation, and, as the wave reaches the boundary, the precursor becomes the local operator ${\rm Tr} \, F^2$.  Finally, we verify that, as expected, the dilaton wave has no time-independent interest contribution.

\subsection{Results in $AdS_5$}

\subsubsection{Bulk Wave as Boundary Squeezed State}

We take the bulk space to be $AdS_5 \times S^5$, although, as nothing will depend on the $S^5$, we will suppress it.  We work in global $AdS_5$ with radius R and, using dimensionless coordinates $B = (\tau, \rho, \Omega)$, where $\rho \in [0,\pi/2)$, with metric
\begin{equation}
ds^2 = \frac{R^2}{\cos^2{\rho}} 
\left(-d\tau^2 + d\rho^2 + \sin^2{\rho} \, d\Omega^2\right).
\end{equation}
In the region near the conformal boundary, we use dimensionful Poincar\'{e}
coordinates $(t, z, \vec{x})$ as an approximation to the above global
coordinates, such that $\tau - \pi/2 \approx t/R$, $\pi/2 - \rho \approx
z/R$, and $\Omega \approx \vec{x}/R$ and $t, r \ll R$.  In this
approximation, the metric near the boundary is
\begin{equation}
ds^2 = \frac{R^2}{z^2} \left(-dt^2 + dz^2 + d\vec{x}^2 \right).
\end{equation}
The conformal boundary is $S^3 \times R$ with coordinates $b = (\tau, \Omega)$ and dimensionless metric
\begin{equation}
\label{boundarymetric}
ds^2 = -d\tau^2 + d \Omega^2\ .
\end{equation}
For distances small compared to $R$, using the change of variables above, the boundary is approximately Minkowski space with dimensionless coordinates $(t/R,\vec{x}/R)$.

\begin{figure}[ht]
\centering
\includegraphics{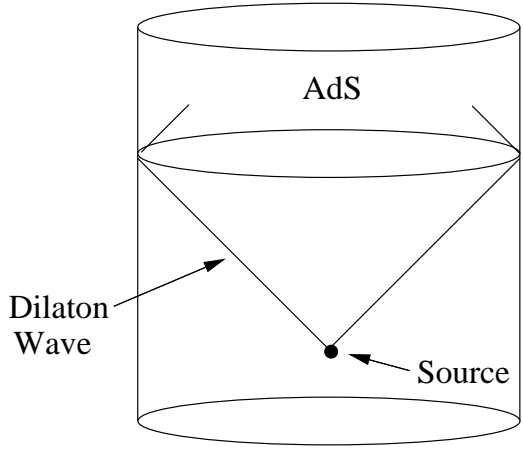}
\caption{\label{Wave} A point-like dilaton source at $\tau=0$ creates an outgoing wave which reaches the boundary at $\tau = \pi/2$ and then reflects back toward the interior.}
\end{figure}

The system we propose to describe, illustrated in Fig.~\ref{Wave}, consists of an outgoing spherical dilaton wave generated at the origin of AdS at $\tau=0$ which reaches the boundary at $\tau=\pi/2$ and reflects back toward the origin.  Prior to the creation of the wave, the bulk is in the vacuum state.  According to AdS/CFT, the boundary state provides an alternative description of this process.  Local boundary observables measured before $\tau=\pi/2$, when the wave reaches the boundary, yield vacuous results.  However, certain nonlocal operators, the precursors, can distinguish between the vacuum and the non-trivial state corresponding to the bulk wave.

We work in the limit that the dilaton is a free, massless bulk field $\phi(B)$ which is
excited at $\tau=0$ by an idealized point source $J(B) = j \delta(\tau)
\delta(\rho)$.  We choose an interaction picture in which the interaction Hamiltonian $H_s(\tau) = \int d\rho \, d\Omega \, J(B) \phi(B)$ generates time evolution for the state. The resulting bulk state is
\begin{eqnarray} 
|j(\tau) \rangle_B  & =  & e^{i \int^\tau_{-\infty} du \, H_s(u)} |0\rangle_B  \nonumber \\
                   \label{fullbulkstate}
		& = & e^{ij \phi(0)}  |0\rangle_B  \: \; \mbox{for} \; \tau >0 \\
		\label{bulkstate}
		& \approx & \lbrace 1 + ij \phi(0) \rbrace |0\rangle_B
\end{eqnarray}
where the subscript $B$ denotes states in the bulk Hilbert space. In (\ref{bulkstate}), we take the source $J(B)$ to be weak and work to linear order in $j$.\footnote{To be precise, if the source, rather than being concentrated at a point, is smeared over a small distance $d \ll R$, we can choose the dimensionful amplitude $j$ such that $\frac{j^2}{d^3} \ll 1$ and then the exponential in (\ref{fullbulkstate}) can be expanded consistently.}

Because $\phi$ is a free field, it can be expanded in the mode sum for scalars in $AdS_{d+1}$ \cite{BGL}:
\begin{equation}
\label{phiexpansion}
\phi(B) = \sum_{n,\ell,\vec{m}} C_{n\ell} \, (\sin{\rho})^{\ell} (\cos{\rho})^{2h_+} P_n^{(\ell+1,\nu)}(\cos{2\rho}) \left\lbrace e^{-i\omega_{n\ell}\tau} Y_{\ell\vec{m}}(\Omega) b_{n\ell\vec{m}} + h.c. \right\rbrace
\end{equation}
where, in $AdS_5$ for massless scalars, the various constants are given by $\nu=2$, $h_+=2$, $\omega_{n\ell}=2n+\ell+2h_+$ and
\begin{equation}
C_{nl} = \left(\frac{\Gamma(n+1)\Gamma(n+\ell+2h_+)}{\Gamma(n+\nu+1)\Gamma(n+\ell+2)} \right)^{\frac{1}{2}} 
\end{equation}
and where $b_{n\ell\vec{m}}^\dagger$ and $b_{n\ell\vec{m}}$ are the bulk creation and annihilation operators.

The bulk field $\phi(B)$ is dual by AdS/CFT to a nonlocal operator $\hat{\phi}(B)$ in the boundary Hilbert space.  For such a bulk field, there is a corresponding local operator $O(b)$, with $b = (\tau,\Omega)$, in the boundary field theory with conformal dimension $\Delta = 2h_+$ given by
\begin{equation}
\label{Ofromphi}
O(b) = \lim_{\rho \to \pi/2} (\cos{\rho})^{-\Delta} \hat{\phi}(b, \rho).
\end{equation}
We can write $\hat{\phi}(B)$ in terms of local boundary operators by $\hat{\phi}(B) = \int db' \, M(B;b') O(b')$ using the transfer matrix $M$ which is given by the mode sum \cite{BGL}
\begin{eqnarray}
\label{Mgeneral}
M(B;b') & = & \sum_{n\ell\vec{m}} \frac{2 (-1)^n \Gamma(n+1)}{2\pi\Gamma(n+3)} (\sin{\rho})^\ell (\cos{\rho})^{2h_+} P^{(\ell+1,2)}_n(\cos{2\rho}) \nonumber \\
&&\hspace{1.5cm}{}\times \left\lbrace e^{-i\omega_{n\ell}(\tau-\tau')}Y_{\ell\vec{m}}(\Omega) Y^*_{\ell\vec{m}}(\Omega') + c.c. \right\rbrace.
\end{eqnarray}

The bulk and boundary Hilbert spaces correspond via AdS/CFT.  We will use a subscript $\partial$ to distinguish boundary states.  The boundary state which is dual to the bulk state (\ref{fullbulkstate}) is, for
$\tau>0$, 
\begin{eqnarray}
|j \rangle_\partial 
\label{boundarystatephi}
& = & e^{ij \hat{\phi}(0)} |0 \rangle_\partial \\
\label{boundarystateO}
& = & e^{ij \int db \, M(0;b) O(b)} |0 \rangle_\partial. 
\end{eqnarray}

The boundary operator $O$  dual to the dilaton is ${\rm Tr} \, F^2$, the square of the gauge field strength.  For the purposes of constructing a simple toy model of the precursors, we will neglect both interactions and the index structure of the fields on the boundary. 
While we expect the indices merely complicate the formulae without substantive effect, ignoring strong interactions may qualitatively affect the results.  In particular, we consider a free, massless canonical scalar $\psi(b)$ on the boundary which will play the role of the gauge potential $A$ and which, in Poincar\'{e} coordinates, can be expanded in modes as
\begin{equation}
\label{psiexpansion}
\psi(b) = \int \frac{d\vec{k}}{(2\pi)^3} \, \frac{1}{\sqrt{2\omega_k}}
\left\lbrace e^{-i\omega_k t + i\vec{k} \cdot \vec{x}} a_{\vec{k}} + h.c. \right\rbrace
\end{equation}
where the energy $\omega_k^2 = \vec{k}^2$ and $a_{\vec{k}}^\dagger$ and
$a_{\vec{k}}$ are boundary creation and annihilation operators.  
Writing $O(b) = CR^4 \partial_\mu \psi(b) \, \partial^\mu \psi(b)$ with $C$ a normalization constant to be determined, and taking the product to be normal ordered, $O$ now models ${\rm Tr} \, F^2$ as desired.

We now proceed to write the boundary state (\ref{boundarystateO}) in a
simple form.  First, we set $B=0$ in (\ref{Mgeneral}) to obtain
\begin{equation}
M(0;b) =  \frac{1}{2\pi^3} \sum_{s=2} \frac{(-1)^s}{s}  \left\lbrace e^{2is\tau} + c.c. \right\rbrace
\end{equation}
where $s=n+2$ and the spherical harmonics are normalized such that $|Y_{0\vec{0}}|^2 = \mbox{Vol}(S^3)^{-1} = (2\pi^2)^{-1}$.  In order to write $M(0;b)$ in the approximate boundary coordinates, we take $\tau - \pi/2 \approx t/R$ where $t \ll R$.  As we will consider lengths small compared with the AdS radius $R$, the dimensionful momentum $q = s/R$, quantized in units of $1/R$, is approximately continuous, and
\begin{equation}
\label{Mapprox}
M(0;b) \approx \frac{1}{2\pi^3} \int_0^{\infty} dq \, \frac{1}{q} \left\lbrace e^{2iqt} + c.c. \right\rbrace\ .
\end{equation}
We plug this expression (\ref{Mapprox}) for $M(0;b)$ and the mode expansion of $O(b)$ into the formula for the boundary state (\ref{boundarystateO}) expanded to linear order in $j$, yielding
\begin{equation}
\label{squeezedstate}
|j \rangle_\partial \approx \left\lbrace 1+\frac{ijC}{2\pi^2} \int \frac{d\vec{k}}{(2\pi)^3} a^\dagger_{\vec{k}} a^\dagger_{-\vec{k}} \right\rbrace |0 \rangle_\partial \ .
\end{equation}
 
This calculation of the squeezed state can be seen as a more concrete derivation of the state (\ref{PSTstate}) postulated in \cite{PST}.  Choosing $F(\vec{k})= \frac{ijC}{\pi^2}$ gives a state identical to (\ref{squeezedstate}) and corresponds to specifying a point source at the origin.\footnote{One can also make a similar comparison with the squeezed state in \cite{SuTo} ({\it c.f.} eq. 2.5).  That state matches (\ref{squeezedstate}) if we choose $f(t)= M(0;b)$.}

\subsubsection{Detecting the Bulk Wave using Precursors}

We would like to perform a measurement in the boundary theory which will be
sensitive to the non-vacuum nature of the state.  Prior to $\tau=\pi/2,
t=0$, when the dilaton wave in the bulk hits the boundary, $n$-point
correlation functions of $O(b)$ in the boundary state
(\ref{boundarystatephi}) give vacuous results by standard field theory
causality and (\ref{Ofromphi}) as follows: 
\begin{eqnarray}
\langle j| O(b_1)...O(b_n) |j \rangle_\partial 
&\sim&  \langle  e^{-ij \hat{\phi}(0)} O(b_1)...O(b_n)  e^{ij \hat{\phi}(0)} \rangle_\partial  \nonumber \\
&\sim& \langle e^{-ij \phi(0)}\phi(b_1,\pi/2)...\phi(b_n,\pi/2) e^{-ij \phi(0)} \rangle_B \nonumber \\
&\sim& \langle \phi(b_1,\pi/2)...\phi(b_n,\pi/2) \rangle_B \nonumber \\
&\sim&  \langle 0|O(b_1)...O(b_n) |0\rangle_\partial  \label{ocommutator}
\end{eqnarray}
where, because each $(b_i,\pi/2)$ is outside the lightcone of the source at the origin, i.e.\ $\tau_i<\pi/2$, we have used $[\phi(0), \phi(b_i,\pi/2)] = 0$ in the third line.  As expected, no product of the analog of local gauge invariant operators is sensitive to the state before the time the wave reaches the boundary.

Instead, we will measure a spacelike two-point function of $\psi(b)$ in the squeezed state (\ref{squeezedstate}) to first order in $j$,
\begin{equation}
\label{psipsia}
\langle j| \psi(b_1) \psi(b_2) |j \rangle_\partial = \frac{jC}{\pi^2}\, \mbox{Im} \,  \int \frac{d\vec{k}}{(2\pi)^3} \, \langle a_{\vec{k}} a_{-\vec{k}} \psi(b_1) \psi(b_2) \rangle_\partial 
\end{equation}
where we have used the hermiticity of $\psi$ to write (\ref{psipsia}) as an imaginary part of a single term.  This is {\it not} analogous to a product of local gauge-invariant operators, since $\psi$ is a model for the gauge potential $A$, which is not gauge invariant.  We will discuss the gauge theory interpretation of these operators shortly. After plugging in the expansion for $\psi$, (\ref{psiexpansion}), and simplifying, we have
\begin{equation}
\langle j| \psi(b_1) \psi(b_2) |j \rangle_\partial = \frac{jC}{\pi^2} \, \mbox{Im} \int \frac{d\vec{k}}{(2\pi)^3} \, \frac{1}{2\omega_k} e^{i\omega_{k}(t_1+t_2)}  \left\lbrace e^{- i\vec{k} \cdot (\vec{x_1} - \vec{x_2})} + c.c. \right\rbrace \ .  \\ 
\end{equation}
We now perform the integral over $\vec{k}$ using spherical coordinates and $\omega_{\vec{k}} = |\vec{k}|$.  Letting $r_{12}=|\vec{x_1}-\vec{x_2}|$ and inserting $i\epsilon$'s to ensure convergence of the integral over $k$, we obtain
\begin{eqnarray}
_\partial\langle j| \psi(b_1) \psi(b_2) |j \rangle_\partial 
&=&  \frac{-jC}{4\pi^4 r_{12}} \, \mbox{Im} \, \left\lbrace \frac{1}{t_1+ t_2-r_{12}+i\epsilon} - \frac{1}{t_1+ t_2 + r_{12}+i\epsilon} \right\rbrace \nonumber \\
\label{psipsi}
&=&  \frac{jC}{4\pi^3 r_{12}} \, \left\lbrace \delta(t_1+ t_2-r_{12}) -
\delta(t_1+ t_2 + r_{12}) \right\rbrace\ .
\end{eqnarray}

In our toy model, this boundary measurement detects the presence of the dilaton wave at $t<0$, before measurements using $O$ could do so.  However, in our model $\psi$ represents the gauge potential $A$ which is gauge dependent and cannot physically be measured.  We might
instead measure $\partial_\mu\psi$, which is more closely analogous to the field strength $F= dA$.

\begin{figure}[ht]
\centering
\includegraphics{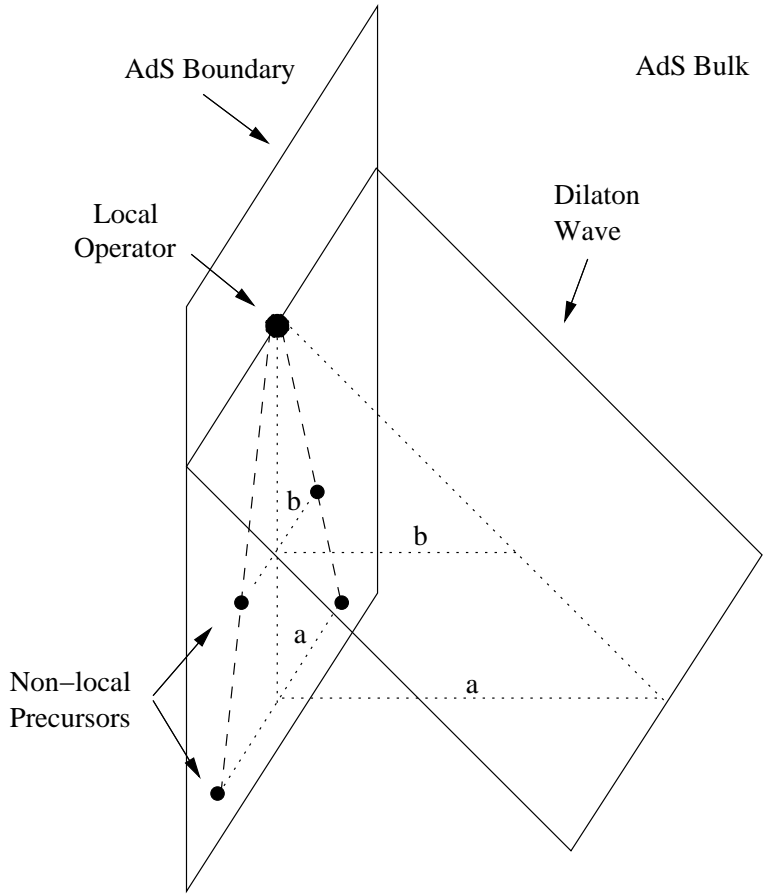}
\caption{\label{WaveAndPrecursors} The size of the precursor with non-zero expectation value decreases as the dilaton wavefront approaches the boundary in accordance with the UV/IR relation.}
\end{figure}

If we now take two derivatives of (\ref{psipsi}) and choose $b_1=(t,\vec{x}-\vec{a})$ and $b_2=(t,\vec{x}+\vec{a})$, we have the simple form
\begin{equation}
\label{precursormeasure}
_\partial\langle j| \partial_{\mu} \psi(b_1) \partial^\mu \psi(b_2) |j \rangle_\partial 
= \frac{jC}{2\pi^4 a} \, \mbox{Im} \, \left\lbrace \frac{1}{(2(t-a)+i\epsilon)^3} - \frac{1}{(2(t+a)+i\epsilon)^3} \right\rbrace.
\end{equation}
The correlation function is proportional to derivatives of $\delta(t-a)-\delta(t+a)$.  The bulk dilaton wave reaches the boundary at $\tau=\pi/2$ or $t=0$, so the wave's distance from the boundary (either before or after collision) is $|t|$.  For a given $a$, a spike is measured when the wave is a distance $a$ from the boundary, as shown in Fig.~\ref{WaveAndPrecursors}.   We identify the nonlocal operator 
\begin{equation}
\label{precursor}
\partial_{\mu}\psi(t,\vec{x}-\vec{a}) \partial^{\mu}\psi(t,\vec{x}+\vec{a})
\end{equation}
as a precursor, which unlike local operators, can distinguish the squeezed state from the vacuum before $t=0$ and thereby detect excitations in the bulk away from the boundary.  

To check the consistency of our model, we let $b_1=b_2=b=(t,\vec{x})$ and multiply by $CR^4$ in (\ref{precursormeasure}) leaving the expectation value of $O(b)$.
\begin{eqnarray}
_\partial\langle j| O(b) |j \rangle_\partial 
&=& \lim_{a \to 0} \frac{jC^2R^4}{2\pi^4 a} \, \mbox{Im} \,  \left\lbrace \frac{1}{(2(t-a)+i\epsilon)^3} - \frac{1}{2(t+a)+i\epsilon)^3} \right\rbrace \nonumber\\
&=& \frac{3jC^2R^4}{8 \pi^4} \, \mbox{Im} \,  \frac{1}{(t+i\epsilon)^4}  
\end{eqnarray}
The imaginary part vanishes for $t \not= \pi/2$, and the result is can be compared to that obtained by direct bulk calculation using the bulk/boundary propagator from \cite{GiddingsFSS}:
\begin{eqnarray}
\lim_{\rho \to \pi/2} (\cos{\rho})^{-2h_+}  {}_B\langle j| \phi(b,\rho) |j \rangle_B 
&=& 2j \, \mbox{Im} \, K_{B\partial}(b,0) \nonumber \\
&\approx& \frac{3jR^4}{16 \pi^2} \mbox{Im} \,  \frac{1}{(t+i\epsilon)^4}\ .
\end{eqnarray}
We find the normalization $C^2 = \pi^2/2$, and verify the model correctly reproduces the prescription of \cite{GKP,Witten}.

As a final check, we investigate whether $O$ has a second-order interest term.  A simple calculation using the squeezed state (\ref{squeezedstate}) and the expansion of $O$ gives
\begin{eqnarray}
_\partial\langle j| O(b) |j \rangle_\partial
&=& {\mathcal O}(j) + \frac{j^2}{\pi^2} \int \, \frac{d\vec{k}}{(2\pi)^3} \frac{\omega_k ^2-\vec{k}^2}{\omega_k} \nonumber \\
&=& {\mathcal O}(j) + 0 \ .
\end{eqnarray}
The boundary stress-energy tensor $T_{\mu \nu}(b) \sim \partial_{\mu} \psi(b) \partial_{\nu} \psi(b)$ has time-independent contributions of order $j^2$ in the squeezed state (\ref{squeezedstate}), which, as in \cite{PST}, can be identified as the interest.   In contrast, the second order contribution to $O(b)$ vanishes.  Because $O$ corresponds to the asymptotic dilaton rather than graviton, we would not expect such a non-linear interest-like correction anyway.  The ability of this toy model, like that of \cite{PST}, to capture aspects of gravitational back-reaction is quite surprising.  We would expect details of the non-linear bulk physics to be beyond the applicability of a free-field model.  To fully reproduce truly gravitational phenomena such as black holes, the strongly interacting SYM is likely required.

\subsection{Results in $AdS_3$}

We can similarly model the precursors of $AdS_3$ and its dual two-dimensional CFT.  The reduction in dimensions simplifies the calculation.  Because the boundary $S^1 \times R$ is flat, we can use the exact geometry instead of an approximation valid near a patch of the boundary.  As a result, the periodicity of the system is explicit and the region far from the boundary is available for investigation.

The outgoing wave in $AdS_3$ can be detected using a boundary two-point function as in  $AdS_5$.  The calculations are similar, and we will merely outline the differences before stating the results.   

We take the bulk space to be $AdS_3 \times S^3 \times T^4$, although, as before, only the $AdS_3$ factor concerns us.  The metric of global $AdS_3$ in dimensionless coordinates $B = (\tau, \rho, \theta)$ is  
\begin{equation}
ds^2 = \frac{R^2}{\cos^2{\rho}} \left(-d\tau^2 + d\rho^2 + \sin^2{\rho} \, d\theta^2 \right).
\end{equation}
Using the coordinates $b = (\tau, \theta)$, the dimensionless metric on the boundary is simply
\begin{equation}
ds^2 = -d\tau^2  +  d\theta^2
\end{equation}
The mode expansion for the massless bulk field $\phi(B)$ is a modification of (\ref{phiexpansion}):
\begin{equation}
\phi(B) = \frac{1}{\sqrt{2 \pi}} \sum_{n,\ell} \left( \frac{n+|\ell|+1}{n+1} \right)^{\frac{1}{2}} (\sin{\rho})^{|\ell|} (\cos{\rho})^2 P_n^{(|\ell|,1)}(\cos{2\rho}) \left\lbrace e^{-i\omega_{n\ell}\tau + i\ell\theta} b_{n\ell} + h.c. \right\rbrace
\end{equation}
with $\nu=1$, $h_+=1$, $\omega_{n\ell}=2(n+|\ell|+1)$, and where $\ell$ ranges from $-\infty$ to $\infty$.   Similarly, the expansion (\ref{psiexpansion}) of the boundary field $\psi(b)$ becomes
\begin{equation}
\psi(b) = \frac{1}{2\pi} \sum_\ell \frac{1}{\sqrt{2\omega_\ell}} \left\lbrace e^{-i\omega_\ell \tau + i\ell\theta} a_\ell + h.c. \right\rbrace
\end{equation}
where the energy $\omega_\ell = |\ell|$.  The transfer matrix $M(0;b)$ (\ref{Mgeneral}) for $AdS_3$ is modified to
\begin{equation}
M(0;b)  = -\frac{1}{(2\pi)^2} \sum_{k=1} \frac{(-1)^k}{k}  \left\lbrace e^{2ik\tau} + c.c. \right\rbrace
\end{equation}
and, by writing $O(b) = C\partial_\mu \psi(b) \partial^\mu \psi(b)$, the state for $\tau > 0$ becomes
\begin{equation}
|j \rangle_\partial \approx \left\lbrace 1-\frac{ijC}{2\pi^2} \sum_{k=1} (-1)^k a^\dagger_k a^\dagger_{-k} \right\rbrace |0 \rangle_\partial.
\end{equation}

Following the steps of the calculation in $AdS_5$, we evaluate a two-point function of $\psi$ to first order in the source $j$:
\begin{equation}
\langle j| \psi(b_1) \psi(b_2) |j \rangle_\partial = \frac{jC}{2\pi^2} \, \mbox{Im} \, \left\lbrace \log \left( 1+e^{i(\tau_1+ \tau_2 -\theta_1+ \theta_2+ i\epsilon)} \right) +  \log \left(1+e^{i(\tau_1 + \tau_2 - \theta_1 + \theta_2+ i\epsilon)} \right) \right\rbrace\ .
\end{equation}
Again, we differentiate each field and set $b_1=(\tau,\theta-a)$ and $b_2=(\tau,\theta+a)$ to obtain the $AdS_3$ analog to (\ref{precursormeasure}),
\begin{equation}
\langle j| \partial_{\mu} \psi(b_1) \partial^\mu \psi(b_2) |j \rangle_\partial = \frac{jC}{4\pi^2} \, \mbox{Im} \, \left\lbrace \frac{1}{\cos^2(\tau+a+ i\epsilon)} + \frac{1}{\cos^2(\tau-a+ i\epsilon)} \right\rbrace 
\end{equation}
As before, we find the correlation function is zero except when the wave is
a distance $a$ from the boundary.  At that time, $|\tau -
\pi(n+\frac{1}{2})| = a$ for any $n \in Z^+$, the resulting delta function
signals the wave's detection.  The periodicity of the result under $\tau
\to \tau + \pi$ corresponds to the wave periodically expanding to the
boundary and reconverging to the origin.  By using the full global
geometry, it becomes apparent that the precursor actually detects the wave
deep in the AdS bulk and our $AdS_5$ result (\ref{precursormeasure}) is not
an artifact of restricting to the near-boundary region.

The expectation value of $O(b)$, obtained by setting $b_1=b_2=b=(t,\theta)$ and multiplying by $C$, matches with the corresponding bulk calculation \cite{GiddingsFSS} for $C^2= \pi$
\begin{eqnarray}
_\partial\langle j| O(b) |j \rangle_\partial 
&=& \frac{jC^2}{4\pi^2} \, \mbox{Im} \, \frac{1}{\cos^2{(\tau + i\epsilon)}}  \nonumber \\
&=& \frac{j}{4\pi} \, \mbox{Im} \, \frac{1}{\cos^2{(\tau + i\epsilon)}}  \nonumber \\
&=& -2j \, \mbox{Im} \, K_{B\partial}(b,0) . 
\end{eqnarray}

\subsection{Successes and Limitations}

The proposed toy model is admittedly oversimplified, yet it manages to
retain those essential properties we expect precursors to exhibit 
in the exact theory.  Excitations deep in the AdS bulk are not manifest in the local
operators of the boundary SYM.  The precursors' defining feature is the ability
to access these hidden degrees of freedom.  Here we have illustrated a
model of a boundary state (\ref{squeezedstate}) which to local operators
appears to be the vacuum, but contains non-trivial correlations measured by
a precursor (\ref{precursormeasure}).

Almost certainly, precursors and other probes of the interior of AdS are
extended, nonlocal objects in the boundary SYM as dictated by the
holographic UV/IR correspondence.  While ultraviolet portions of the field
theory correspond to the infrared boundary asymptotics of AdS, long-distance length
aspects encode information regarding regions of AdS whose
distance from the boundary is directly related to their size.  In our model,
the information that the wave is a distance $a$ from the boundary is
contained in the non-zero correlator of fields separated by a distance $2a$
on the boundary.  This type of UV/IR relation was seen in \cite{PST, SuTo},
and in many other contexts.

Despite these successes, not surprisingly our model fails to capture all the aspects of the
correspondence between string theory in AdS and its CFT dual.  The precursor (\ref{precursor}) gives some information about the bulk, but doesn't precisely correspond to $\phi(t,a,\vec{x})$ or to $\phi$ anywhere else in the bulk.  For instance, when $|t|<a$, $\langle j|\phi(B)|j\rangle \not= 0$, while the precursor (\ref{precursormeasure}) is identically zero. This is hardly surprising as the direct holographic dual to $\phi(B)$, at least in this free-field model, is $\hat{\phi}(B) = \int db' \, M(B;b') O(b')$.  However, not only does $\hat{\phi}(B)$ lack a simple SYM analog, but it is completely nonlocal in both space and time.

As noted previously in section 2.1, our model uses an unjustified
free-field approximation.  In the 't Hooft limit, where string theory is
well described by supergravity, the coupling $g^2_{YM}N$ of the boundary
SYM is large, yet we employ just the opposite, a non-interacting model.
While some attempts to justify this approach were presented in \cite{PST},
we are content to recognize conclusions based upon such a model are
suggestive of the results of more accurate calculations.  

Given these objections, that our simple free-field model works at all is somewhat surprising.  One reason why lies in the fact that while $O$ has a simple interpretation in the bulk field theory, neither our precursor (\ref{precursor}) nor the fields $\psi$ from which it's composed correspond
to identifiable field theory objects.  Although there are many reasons to distrust a free boundary field theory, the most important flaw in our model may be in the bulk.  When bulk interactions are added, the correspondence (\ref{Ofromphi}) can not be inverted so easily to give bulk fields in terms of boundary operators.  Furthermore, for the bulk theory to have any holographic description, it must necessarily nonlocal.  Both our simple bulk theory and weakly interacting supergravity are local field theories.  An improved calculation in interacting SYM with nonlocal precursors would require string theory for a complete bulk description.  We next turn to a discussion of how the precursors might be identified in this full ${\cal N}=4$ $SU(N)$ gauge theory.

\sect{Precursors and Bulk Locality}

The obvious question is what light does our toy model shed on the possible
identity of the precursors in AdS/CFT.  Our model does demonstrate some
points of principle.  The first is that it is possible to encode
information in a boundary state such that it cannot be measured by a local
observation, but instead requires measuring a nonlocal product of
operators, as advocated in \cite{PST}.  Information is encoded this way by
virtue of the boundary state being a squeezed state of a specific type.  
The second is that the nonlocality,
as exhibited by the separation of the operators, acts in accord with
the expectations of the UV/IR correspondence: larger distances on the
boundary correspond to events deeper into the bulk.

However, while serving as a guide, our model is not yet sufficient to
precisely identify the precursors in the true gauge theory.  Of course,
Wilson loops are nonlocal operators, and thus natural candidates.
However, since the general observables in a nonabelian gauge theory are
Wilson loops, this is not sufficiently specific.  Susskind and Toumbas
\cite{SuTo} suggested that measurement of large, smooth Wilson loops is
sufficient to detect information deep in the bulk.  But \cite{GiLi}
demonstrated a flaw in the calculation purporting to demonstrate this.
What is desired is a more specific description of exactly what kind of
Wilson loop measurements allow one to access information outside the bulk
lightcone.

In seeking the identity of the precursors, we should take as many clues as
possible from the bulk viewpoint.  In particular, if the AdS/CFT correspondence indeed provides an accurate representation of physics that is approximately local in the bulk,  on scales large as compared to the string distance, then the precursors should respect the constraints expected from this approximate bulk locality.  We assume that the boundary observables that correspond to the precursors are some appropriate limit of bulk operators or observables.

\begin{figure}[ht]
\centering
\includegraphics{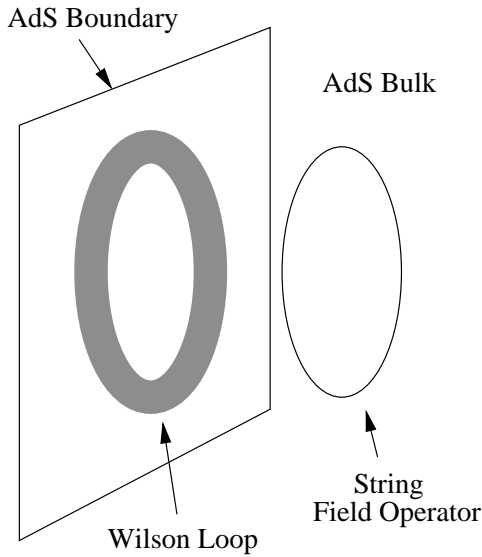}
\caption{\label{LoopAndString} A Wilson loop $W[C]$ in the boundary theory with UV cutoff $\delta$ corresponds  in the bulk to a string field operator $\Phi[\rho=\pi/2-\delta, C]$.}
\end{figure}

For example, according to the proposal of \cite{MaldW, ReYe} Wilson loops of the form
\begin{equation}
W[C] = \frac{1}{N}{\rm Tr} \, \mathcal{P} \, e^{i \oint d\sigma' \left\lbrace {\dot x}^\mu(\sigma') A_{\mu} + {\dot y^i(\sigma')} \phi_i \right\rbrace} \ ,
\end{equation}
where $A_{\mu}$ is the $SU(N)$ gauge field, $\phi_i$ are the six scalars, and $y^i$ is the path of the loop on $S^5$ such that ${\dot y}^2 = {\dot x}^2$, are dual to the boundaries of string world sheets.  As was argued by two of the authors in \cite{GiLi} (see also \cite{Bhat} for further discussion), this indicates that observation of a Wilson loop corresponds to an appropriate limit of a measurement of the string field operator $\Phi[x(\sigma)]$ (Fig.~\ref{LoopAndString}),
\begin{equation}
W[C] \leftrightarrow \lim_{x(\sigma)\rightarrow C}
Z[x(\sigma)] \Phi[x(\sigma)]\ ,
\end{equation}
where $Z[x(\sigma)]$ is a normalization factor analogous to the factor $\cos^{-\Delta}{\rho}$ (see \ref{Ofromphi}) for local operators in the prescription of \cite{GKP,Witten}.  Here we are being somewhat schematic, as a general definition of the string field operator is fraught with difficulties, gauge and otherwise (the former can in some cases be avoided by working in lightcone gauge).  This means that we are interested in a limit of the expectation value 
\begin{equation}
\label{stringvev}
\langle j| \Phi[x(\sigma)] | j\rangle_B 
\end{equation} 
of the string field operator.  The question is, from the bulk perspective, under what circumstances this expectation value would be expected to be non-vanishing outside the lightcone of the source.

Ref.~\cite{GiLi} formulated a {\it locality bound} which gives a general criterion:  bulk locality should fail when we probe field components with appreciable overlap with states of high enough energy to imply locality violations, for example by either producing a long string or a black hole that extends over the region containing both the source and the observation point.

It is in practice rather difficult to explore the black hole version of this statement, since it intrinsically refers to non-perturbative gravitational phenomena. There is some more hope of investigating the role of the extended nature of strings in this context.

In particular, if these effects can be exhibited in string perturbation theory, the simplest contribution to (\ref{stringvev}) should come from the string two-point function,
\begin{equation}
\langle 0|
\left[\Phi\left[x(\sigma)\right],\Phi\left[x'(\sigma)\right]\right] 
|0\rangle 
\end{equation}
(compare eq.~\ref{ocommutator}).  Possibly nonlocal effects in this commutator are only poorly understood.  These were discussed in \cite{Lowe,LSU,LPSTU} which suggested that, using interaction lightcone string field theory, such effects could be exhibited, but fell short of a clear gauge invariant statement.  The essential question is what components of the string field expectation value (\ref{stringvev}) we expect to manifest the nonlocal effects.

An argument given in \cite{Santa} states that whatever the precursors are, they cannot be measured by a collection of local observers.  The logic was that the time for a signal to go from the center of AdS to the boundary could be made arbitrarily long by adding matter.  On the other hand, a measurement in the field theory which is made by a collection of local observers need only take as long as it takes to collect the information to a central point.  This time is bounded.  We suggest an alternative resolution to the paradox:  bulk locality is violated.  The reasoning of \cite{Santa} relies on bulk locality in assuming that no excitation can be detected at the boundary until a light ray has time to reach it.  We believe that there is evidence that the measurement of Wilson loops, for example, can detect information nonlocally within the constraints of the locality bound of \cite{GiLi} if the measurements are carried out on a sufficiently fine scale.

Specifically, let us suppose that the boundary gauge theory is cutoff at a scale $\delta$ with respect to the dimensionless boundary metric (\ref{boundarymetric}).  How do we for example measure a spacelike Wilson loop?  Indeed, the authors of ref.~\cite{Beckman:2001ck} argue that Wilson loops are not observables in the traditional sense, in that they can't be measured in a non-demolition measurement:  if a Wilson loop is measured, even eigenstates of the Wilson loop operator are generically disturbed.  However, they can be measured in a demolition experiment.  Specifically, suppose that we want to measure a Wilson loop at a resolution given by $\delta$.  We can prearrange charged particles at separations $\delta$ along the curve to be measured, and at the agreed upon time, each of these particles can be moved to a neighboring location, measuring the phase of the wavefunction.  Since this measurement takes a time $\delta$, we expect that the energy associated to the measurement of each component particle must be of order $1/\delta$, in the same sense that measurement of the value of an ordinary scalar field on a time scale $\Delta t$ involves field involves creation or annihilation of quanta of energy $1/\Delta t$.  If $L$ is the length of the Wilson loop, there are $L/\delta$ such particles, corresponding to a characteristic energy
\begin{equation}
E_{\rm loop} \sim \frac{L}{\delta^2} \label{gaugeen}
\end{equation}
for the measurement.  

Let us discuss the corresponding bulk measurement, which we have argued corresponds to absorbing a large string.  One way of implementing the cutoff with resolution $\delta$ in the field theory is to move the AdS boundary in to $\rho=\pi/2-\delta$. As shown in Fig.~\ref{LoopAndString}, we imagine absorbing a string on this surface.  For a Wilson loop of length $L$, the corresponding string will have proper length $LR/\delta$.  Consider, for example, a maximal length loop that spans the equator, with $L \sim 1$ and proper length $R/\delta$.  The above measurements correspond to measuring the string over a proper time $T\sim R$ and with a resolution size $R$; this is just the statement that physics at the cutoff corresponds to physics in a single AdS radius \cite{SusskindHFSL}.  Thus, the proper energy required to measure the string, thought of as bits of size $R$ and resolved on times $R$, is 
\begin{equation}
E_{\rm meas} \sim \frac{1}{\delta R}\ .
\end{equation}
This is of course exactly what we get by converting (\ref{gaugeen}) to bulk units.  

To check the relationship to the locality bound, let's estimate the energy of a string stretching from the center of AdS to the boundary at $\rho=\pi/2-\delta$.  The Nambu-Goto action takes the form
\begin{equation}
S\sim \frac{1}{\alpha'} \int d\tau \int^{\pi/2 -\delta} d\rho \, \frac{R^2}{\cos^2\rho} \sim \frac{\Delta \tau R^2}{\alpha' \delta}
\end{equation}
In gauge theory units, this corresponds to an energy $E_{\rm string}\sim R^2/\alpha' \delta$.  Comparing this to $E_{\rm loop}$, we see that the energy of the loop measurement is higher than that of the stretched string state if $\delta \stackrel{<}{\sim} (l_s/R)^2$.  This suggests that the measurement should indeed involve energies beyond the saturation of the locality bound, providing a possible rationale for bulk locality violation.

This discussion also makes it clear that the information is contained in the structure of Wilson loops on a fine scale.  If we average over much larger scales, the loop will not saturate the locality bound.  Precursor variables, if they indeed arise in this fashion, are encoded in {\it short-distance} variations of these loops and not in their coarse-grained average.  

A particular kind of Wilson loops that is sensitive to short-distance information is the {\it decorated} Wilson loops, namely Wilson loops with insertions of local operators, {\it e.g.}~the field
strength $F$, at points along the loop.  Our toy calculation is also suggestive of such operators.  Decorated loops correspond to variations of ordinary loops:
\begin{eqnarray}
\frac{\delta}{\delta x^\mu(\sigma)} W[x(\sigma')] 
&=& \frac{\delta}{\delta x^\mu(\sigma)} \frac{1}{N}{\rm Tr} \, \mathcal{P} \, e^{i \oint d\sigma' \left\lbrace {\dot x}^\mu(\sigma') A_{\mu} + {\dot y^i(\sigma')} \phi_i \right\rbrace} \nonumber \\
&=& \frac{i}{N}{\rm Tr} \, \mathcal{P}  \left\lbrace F_{\mu \nu} (x(\sigma)) {\dot x}^\nu + \partial_\mu \phi_i(x(\sigma)) {\dot y^i(\sigma')} \right\rbrace e^{i \oint d\sigma' \left\lbrace {\dot x}^\mu A_{\mu} + {\dot y^i} \phi_i \right\rbrace}
\end{eqnarray}
Via the correspondence between Wilson loops and string fields, we similarly see that these correspond to variations of the string fields at short scales:
\begin{equation}
\frac{\delta}{\delta x^\mu(\sigma)} \Phi[x(\sigma')]\ .
\end{equation}
These comments suggest that decorated loops may be an important part of the story.

\begin{figure}[ht]
\centering
\includegraphics{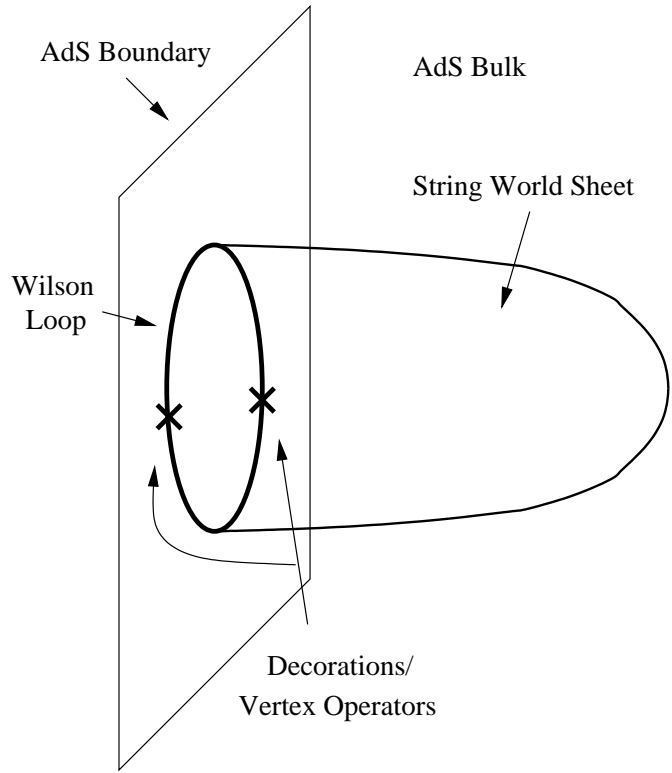}
\caption{\label{Decorations} A decorated Wilson loop corresponds in the bulk to a world sheet with vertex operators at the locations of the decorations.}
\end{figure}

It's also amusing to think about the bulk description of a process corresponding to measuring a decorated Wilson loop.  In a weakly-coupled D-brane description, a Wilson loop corresponds to a string end traversing a curve on the boundary; for a stack of branes we must sum over the different branes on which the string can terminate.  Insertion of an operator such as the field strength $F$ corresponds to absorbing a short string that connects one brane to another.  To understand the corresponding AdS picture, we must answer the question  ``where are the branes?"  Susskind \cite{SusskindHawkingfest} has suggested that they lie at the boundary, since in the UV
\begin{equation}
\langle {\rm Tr} \phi^i \phi^i\rangle \sim \langle r^2\rangle =\infty\ .
\end{equation}
This shows that their mean position is at infinity.
Moreover, large-N factorization
\begin{equation}
\langle {\rm Tr} \phi^i \phi^i   {\rm Tr} \phi^j \phi^j\rangle - \langle{\rm Tr} \phi^i \phi^i\rangle^2=0 
\end{equation}
says that the fluctuations from infinity vanish.  Thus, the AdS picture of a decorated Wilson loop is apparently that of a world sheet terminating on the boundary, with insertion of  string vertex operators corresponding to absorbing a short strings at the AdS boundary, as shown in Fig.~\ref{Decorations}.  This fits well with the fact that the insertion $F$ is a local operator and thus should be supported in the UV.

\sect{Conclusion}

The exact description of the precursors and the corresponding problem of ``decoding the hologram" remain somewhat of a mystery.  We believe we have contributed  towards clarifying the possible identification of the precursors.  Specifically, we have seen that a model for precursors can be explicitly analyzed in which no local boundary observable detects the information contained in the state, but a bilocal operator does.  The corresponding state is a special form of squeezed state.  It is tempting to speculate that this basic picture translates into the strongly coupled gauge theory relevant at large $R$.  We have argued that it is indeed plausible that Wilson loops, if measured on fine enough scales, do indeed detect bulk information ``at a distance," and specifically that this is possible within the constraints of the locality bound:  observation of such a Wilson loop is a high-energy observation, with sufficient energy to create a stretched string and probe the intrinsic nonlocality of string theory.  Decorated loops, which make more explicit the short distance structure of the theory, may in particular play an important role.
However, we have yet to exhibit a calculation that demonstrates that the information is indeed detected by Wilson loops.  Such a calculation is apparently beyond present technology in $AdS_5$, as it requires a more complete understanding of string propagation, beyond the supergravity approximation, in that space.

\bigskip
\bigskip
\centerline{\bf Acknowledgments}
\medskip
The authors would like to thank Andrew Frey, Veronika Hubeny, Matt Kleban, John McGreevy, Joe Polchinski, 
Scott Thomas, Nicolaos Toumbas, Brookie Williams, and especially Lenny Susskind for valuable
conversations.  In addition, we would like to thank Stanford for their hospitality to those of us from Santa Barbara.  This research was supported in part by Department of Energy under contract DE-FG-03-91ER40618, and by the David and Lucile Packard foundation.  SBG would like to acknowledge the hospitality of the Newton Institute, where this work was finalized.
\newpage
\providecommand{\href}[2]{#2}
\begingroup\raggedright

\endgroup


\begin{thebibliography}{10}

\bibitem{tHoo}
G. 't Hooft, ``Dimensional Reduction in Quantum Gravity,'' gr-qc/9310026.

\bibitem{Suss}
L. Susskind, ``The World as a Hologram,'' hep-th/9409089, {\sl J. Math. Phys.} {\bf 36} (1995) 6377.

\bibitem{Mald}
J. Maldacena, ``The Large N Limit of Superconformal Field Theories and Supergravity,'' hep-th/9711200, {\sl Adv. Theor. Math. Phys.} {\bf 2} (1998) 231.

\bibitem{magoo}
O.~Aharony, S.S.~Gubser, J.~Maldacena, H.~Ooguri and Y.~Oz,
``Large N field theories, string theory and gravity,'' hep-th/9905111.

\bibitem{GKP}
S.~S. Gubser, I.~R. Klebanov, and A.~M. Polyakov, ``Gauge theory correlators from noncritical string theory,'' hep-th/9802109, {\sl Phys. Lett.} {\bf B428} (1998) 105.

\bibitem{Witten}
E.~Witten, ``Anti De Sitter Space And Holography,'' hep-th/9802150, {\sl Adv. Theor. Math. Phys.} 2 (1998) 253

\bibitem{GiddingsBSM}
S.~B.~Giddings, ``The boundary S-matrix and the AdS to CFT dictionary,'' hep-th/9903048, Phys.\ Rev.\ Lett.\  {\bf 83}, 2707 (1999).

\bibitem{SuWi}
L.~Susskind and E.~Witten, ``The Holographic Bound in Anti-de Sitter Space,'' hep-th/9805114.

\bibitem{info}
See, {\it e.g.} 
S.B. Giddings, ``The black hole information paradox,'' hep-th/9508151;
 ``Quantum Mechanics of Black Holes,'' hep-th/9412138;
A. Strominger, ``Les Houches Lectures on Black Holes,''
hep-th/9501071, NATO Advanced Study Institute: Les Houches Summer School, 
Session 62: {\sl Fluctuating Geometries in Statistical 
Mechanics and Field Theory}, F. David, P. Ginsparg, and J. Zinn-Justin
(eds.).

\bibitem{LoTh}
D.A.~Lowe and L.~Thorlacius,``AdS/CFT and the Information Paradox,'' hep-th/9903237,
{\sl Phys.Rev.} {\bf D60} (1999) 104012.

\bibitem{Nu}
S.~B.~Giddings and A.~Nudelman, ``Gravitational collapse and its boundary description in AdS,'' hep-th/0112099, JHEP {\bf 0202}, 003 (2002).

\bibitem{PST}
J. Polchinski, L. Susskind, N. Toumbas, ``Negative Energy, Superluminosity and Holography,'' hep-th/9903228, {\sl Phys. Rev.} {\bf D60} (1999) 084006.

\bibitem{SuTo}
L. Susskind and N. Toumbas, ``Wilson Loops as Precursors,'' hep-th/990901, {\sl Phys. Rev.} {\bf D61} (2000) 044001.

\bibitem{MaldW}
J. Maldacena, ``Wilson loops in large N field theories,'' hep-th/9803002, {\sl Phys. Rev. Lett.} {\bf 80} (1998) 4859.

\bibitem{ReYe}
S.J.~Rey and J.T.~Yee, ``Macroscopic strings as heavy quarks: Large-N gauge theory and anti-de Sitter supergravity,'' {\sl Eur. Phys. J.} {\bf C22} (2001) 379, hep-th/9803001.

\bibitem{BCFM}
D. Berenstein, R. Corrado, W. Fischler, and  J. Maldacena, ``The Operator Product Expansion for Wilson Loops and Surfaces in the Large N Limit,'' hep-th/9809188, {\sl Phys. Rev.} {\bf D59} (1999) 105023.

\bibitem{GiLi}
S. Giddings and M. Lippert, ``Precursors, Black Holes, and a Locality Bound,'' hep-th/0103231.

\bibitem{BGL}
V. Balasubramanian , S. B. Giddings, and  A. Lawrence, ``What Do  CFTs Tell Us About Anti-de Sitter Spacetimes?'' hep-th/9902052, {\sl JHEP} {\bf 9903} (1999) 001.

\bibitem{GiddingsFSS}
S.B. Giddings, ``Flat-space scattering and bulk locality in the  AdS/CFT correspondence,'' hep-th/9907129, {\sl Phys. Rev.} {\bf D61} (2000) 106008.

\bibitem{Santa}
M.~Kleban, J.~McGreevy, and S.~Thomas, ``Photon telegram to S. Claus: Consequences for holography and causality in AdS,'' hep-th/0112229.

\bibitem{Lowe}
 D.A. Lowe, ``Causal Properties of Free String Field Theory,'' hep-th/9312107, {\sl Phys. Lett.} {\bf B326} (1994) 223.

\bibitem{LSU}
 D. A. Lowe, L. Susskind, J. Uglum, ``Information Spreading in Interacting String Field Theory,'' hep-th/9402136, {\sl Phys. Lett.} {\bf B327} (1994) 226.

\bibitem{LPSTU}
D. A. Lowe, J. Polchinski, L. Susskind, L. Thorlacius, and J. Uglum, ``Black Hole Complementarity vs. Locality,'' hep-th/9506138, {\sl Phys. Rev.} {\bf D52} (1995) 6997.

\bibitem{Bhat}
S.~Bhattacharya, ``A Short Note On The Wilson-Loop Average And The AdS/CFT Correspondence,'' hep-th/0202088.

\bibitem{Beckman:2001ck}
D.~Beckman, D.~Gottesman, A.~Kitaev and J.~Preskill, ``Measurability of Wilson loop operators,'' hep-th/0110205, Phys.\ Rev.\ D {\bf 65}, 065022 (2002).

\bibitem{SusskindHFSL}
L.~Susskind, ``Holography in the Flat Space Limit,'' hep-th/9901079.

\bibitem{SusskindHawkingfest}
L.~Susskind, ``Twenty Years of Debate with Stephen,'' hep-th/0204027.

\end{thebibliography}
\end{document}